 \documentclass[twocolumn,trackchanges]{aastex62}

\shorttitle{Rotational evolution of PSR J0250+5854}

\begin{document}

\title{Rotational Evolution of The Slowest Radio Pulsar PSR J0250+5854}

\correspondingauthor{ H.Tong}
\email{htong$\_$2005@163.com}


\author[0000-0002-0069-831X]{F.F.,Kou}
\affil{CAS Key Laboratory of FAST, National Astronomical Observatories, Chinese Academy of Sciences, Beijing 100101, China}
\affiliation{Xinjiang Astronomical Observatories, Chinese Academy of Sciences, Urumqi 830011, China}

\author{H.Tong}
\affiliation{School of Physics and Electronic Engineering, Guangzhou University, Guangzhou 510006, China}

\author{R. X., Xu}
\affiliation{School of Physics and State Key Laboratory of Nuclear Physics and Technology, Peking University, Beijing 100871, China}
\affiliation{Kavil Institute for Astronomy and Astrophysics, Peking University, Beijing 100871, China}

\author{X., Zhou}
\affiliation{Xinjiang Astronomical Observatories, Chinese Academy of Sciences, Urumqi 830011, China}
\affiliation{Key Laboratory of Radio Astronomy, Chinese Academy of Sciences}


\begin{abstract}
We apply theoretical spin-down models of magnetospheric evolution and magnetic field decay to simulate the possible evolution of PSR J0250+5854, which is the slowest-spinning radio pulsar detected to date. Considering the alignment of inclination angle in a $3$-D magnetosphere, it is possible that PSR J0250+5854 has a  high magnetic field comparable with magnetars or/and high magnetic field pulsars, if a small inclination angle is considered. Our calculations show that similar long-period pulsars tend to have a relatively low period derivative in this case.  In another case of magnetic field decay, calculations also show a possible connection between PSR J0250+5854 and high dipole-magnetic field magnetars. The evolutionary path indicates a relatively high spin-down rate for  similar long-period pulsars.
\end{abstract}


\section{Introduction} \label{sec:intro}
More than $2600$ pulsars have been observed since the first discovery of pulsar\footnote{https://www.atnf.csiro.au/research/pulsar/psrcat/} \citep{1968_Hewish,2005_Manchester_ATNF}. They are divided into different groups according to their observational properties \citep{2013_Harding_pulsar_groups}. The evolutionary connections between different groups of neutron stars are widely discussed with different ways. The Rotating Radio Transients (RRATs for short) are scattered throughout the range of normal pulsars on the $P$-$\dot{P}$ diagram, but with relatively larger characteristic ages. They are taken as old pulsars due to the extreme nulling behaviour in terms of the radiation \citep{2007_Zhang_RRATs_death}.  The high-energy bursts of young high-magnetic field pulsars look very similar to the emission of magnetars \citep{2008_Gavriil_1846,2016_1119_burst}. Possible evolutionary routes from high-$\dot{P}$ rotation-powered pulsars to magnetars are studied by modifying  braking mechanisms \citep{2011Espinoza,2014_Liu_fbd}. X-ray Dim Isolated Neutron Stars (XDINSs for short) are located just below the magnetars on the $P$-$\dot{P}$ diagram and identified as old magnetars according to the studies of magneto-thermal evolution \citep{2013_Vigan_thermal}.

Our traditional understandings about the relationship between different groups of pulsars may be updated by the discovery of  long-period of pulsars. PSR J0250+5854,  the slowest rotating radio pulsar detected to date \citep{2018_Tan_0250}, its  period and period  derivative are $23.5\, \rm s$  and $ 2.7 \times 10^{-14} \, \rm s/s $, which derives a dipole magnetic field of $5.1\times10^{13}\, \rm G$ at the pole and a characteristic age of $13.7 \, \rm Myr$. 
On the one hand, PSR J0250+5854 is located under the conventional definition of death-line where radio emission is expected to stop, which challenges traditional knowledge about pulsar radio emission. On the other hand, though the spin period of PSR J0250+5854 is similar to that of magnetars or XDINSs, it is hard to make connection between them due to the lack of detection of high-energy emission.

To investigate the link between PSR J0250+5854 and other neutron stars,  we respectively apply spin down models of magnetospheric evolution and magnetic field decay to simulate its  possible rotational evolutions and to explain its unusual observations. Besides, possible observational properties under these two cases are predicted.

\section{Rotational evolution  in the case of magnetospheric evolution} \label{sec:style}
\subsection{Spin-down torque}

It is generally accepted that the neutron star should be surrounded by a plasma filled magnetosphere.  There are particles accelerated in the magnetosphere. The radiation of particles will generate the observed pulsar emission. The out-flowing particle winds will take away the rotational energy to spin down the pulsar. Meanwhile, the presence of plasma in the magnetosphere will also generate a torque to align  the magnetic axis and the rotational axis \citep{1970_Michel_alignment,2014_Philippov_3D}. For an oblique rotator, the evolution equation is \citep{1970_Michel_alignment,2014_Philippov_3D} 
 
\begin{equation}
 I\frac{d \textbf{$\Omega$}}{dt}=\textbf{\textit {K}},
 \label{equ:Omega1}
 \end{equation}
where $I$ is the moment of inertia and a typical value of $10^{45} \rm g \, cm^2$ is taken in the following calculations, $\Omega$ is the angular velocity, and $\textbf{\textit {K}}$ is the torque working on the pulsar. For a spherical system, Equation (\ref{equ:Omega1}) can be expressed as 

\begin{equation}
\label{equ:Omage2}
I \frac{d\Omega}{dt}=K_{\rm spin-down},
\end{equation}

\begin{equation}
\centering
\label{equ:alpha}
I\Omega\frac{d\alpha}{dt}=-K_{\rm alignment},
\end{equation}
where $\alpha$ is the angle between the magnetic axis and the rotational axis (i.e., the inclination angle), $K_{\rm spin-down}$ is the torque to spin down the pulsar, and $K_{\rm alignment}$ is the torque to align the pulsar. 

According to the MHD simulations of pulsar magnetosphere \citep{2006_Spitkovsky_MHD},  the spin-down torque can be expressed as 
\begin{equation}
K_{\rm spin-down}=-k_{0}\frac{\mu^2\Omega^3}{c^3}(\sin^2\alpha+k_{1}),
\label{equ:spin-torque}
\end{equation}
where $\mu=(1/2) BR^3$ is the magnetic dipole moment ($B$ is the polar magnetic field and $R$ is the radius of the pulsar), and $c$ is the speed of light. Generally, $k_{0}$ is a numerical factor, $k_{1}$ represents the effect of out-flowing particles and it is model dependent. Comparisons between different MHD simulations were discussed in \citet{2017_Tong_alignment}. Here, we take $k_{0} \approx k_{1}\approx1$ for simplify.  As pulsar spinning down, the ability of pair production will decrease \citep{1975ApJ...196...51R}. For pulsars near the death line, the effect of death must be considered \citep{2000_Zhang_pulsar_death,2006_Contopoulos_spindown}.  \citet{2006_Contopoulos_spindown} applied the effect of pulsar death to simulate the rotational evolution of pulsars.  We employ their treatment of pulsar death, and Equation (\ref{equ:Omage2}) could be expressed as
\begin{equation}
I\frac{d\Omega}{dt}= 
-\frac{\mu^2}{c^3}\Omega^{3}\left\{ \begin{array}{lr} \sin^{2}\alpha +
 (1-\frac{\Omega_{\rm death}}{\Omega})        & \mbox{if
$\Omega > \Omega_{\rm death}$} \\
\sin^{2}\alpha                                    & \mbox{if
$\Omega \leq \Omega_{\rm death}$}   \end{array} \right.,
\label{equ:Omega3}
\end{equation}
where $\Omega_{\rm death}=2\pi/P_{\rm death}$ is the angular velocity when pulsar dies, and the death period is defined as \citep{2006_Contopoulos_spindown,2012_Tong_SGR0418}
\begin{equation}
\label{equ:pdeath}
P_{\rm death}=0.885\Big( \frac{B}{10^{12}\rm G} \Big)^{1/2} \Big(\frac{V_{\rm gap}}{10^{13}\rm V} \Big)^{-1/2} \rm \, s,
\end{equation}
where $V_{\rm gap}$ is the potential drop of the acceleration gap. Equation (\ref{equ:Omega3}) means that the pulsar is braked by the combination of magneto-dipole radiation and particle wind before its death, but only braked  by the magneto-dipole radiation after its death. Compared with the previous work of \citep{2006_Contopoulos_spindown}, a factor of $\cos^{2} \alpha$ is omitted here. As discussed in \citet{2017_Tong_alignment}, it is a weighting factor between the magnetic-dipole radiation and the magnetospheric particles.  Besides, considering the result of the magnetospheric simulations \citep{2006_Spitkovsky_MHD,2012_Li_mag}, the $\cos^{2} \alpha$ factor may not appear in the particle wind component.

As the two components of the spin-down torque are independent of the inclination angle when $\alpha=0^{\circ}$ and $\alpha=90^{\circ}$, the alignment torque and the spin-down torque can be related as: $K_{\rm alignment}=[K_{\rm spin-down}(0^{\circ})-K_{\rm spin-down}(90^{\circ})]\sin\alpha \cos\alpha$ \citep{2014_Philippov_3D}. Hence, Equation (\ref{equ:alpha}) could be expressed as
\begin{equation}
I\Omega\frac{d\alpha}{dt}=-\frac{\mu^2}{c^3}\Omega^{3}\sin \alpha \cos \alpha.
\label{equ:alpha2}
\end{equation}

\subsection{Spin-down of PSR J0250+5854}
As PSR J0250+5854 is located just below the death line on the $P$-$\dot{P}$ diagram, its radio emission may tend to stop, and its period may be much close to the death period, $P_{\rm obs} \le P_{\rm death}$. An  magnetic field of $B \geq 1.276  \times 10^{12}P^{2}_{\rm obs}(V_{\rm gap}/10^{13} \, \rm V) \, \rm G$ can be calculated by Equation (\ref{equ:pdeath}). Hence,  an  upper limit on  the inclination angle of $\alpha=3.4^{\circ}(V_{\rm gap}/10^{13} \rm \, V)^{-1}$ can be calculated by Equation (\ref{equ:Omega3}). Generally, $V_{\rm gap}=10^{13}\, \rm V$ is taken for normal pulsars \citep{2006_Contopoulos_spindown}, and the corresponding minimum magnetic field and maximum inclination angle are $7.06\times 10^{14} \, \rm G$ and $3.4^{\circ}$, respectively. Because PSR J0250+5854 is still a radio loud pulsar, an inclination angle of $2^{\circ}$ is assumed.  The  magnetic field  could be calculated by Equations (\ref{equ:Omega3}) and (\ref{equ:pdeath}), which is about $7.1\times10^{14} \, \rm G$\footnote{For pulsars near the death line, the effect of particle wind would decrease. Meanwhile, if the inclination angle is very small, their contributions on the braking torque from magneto-dipole radiation and particle wind are comparable. Hence, the changes in magnetic field are small with different inclination angles.}.

\begin{figure}[htp]
\plotone{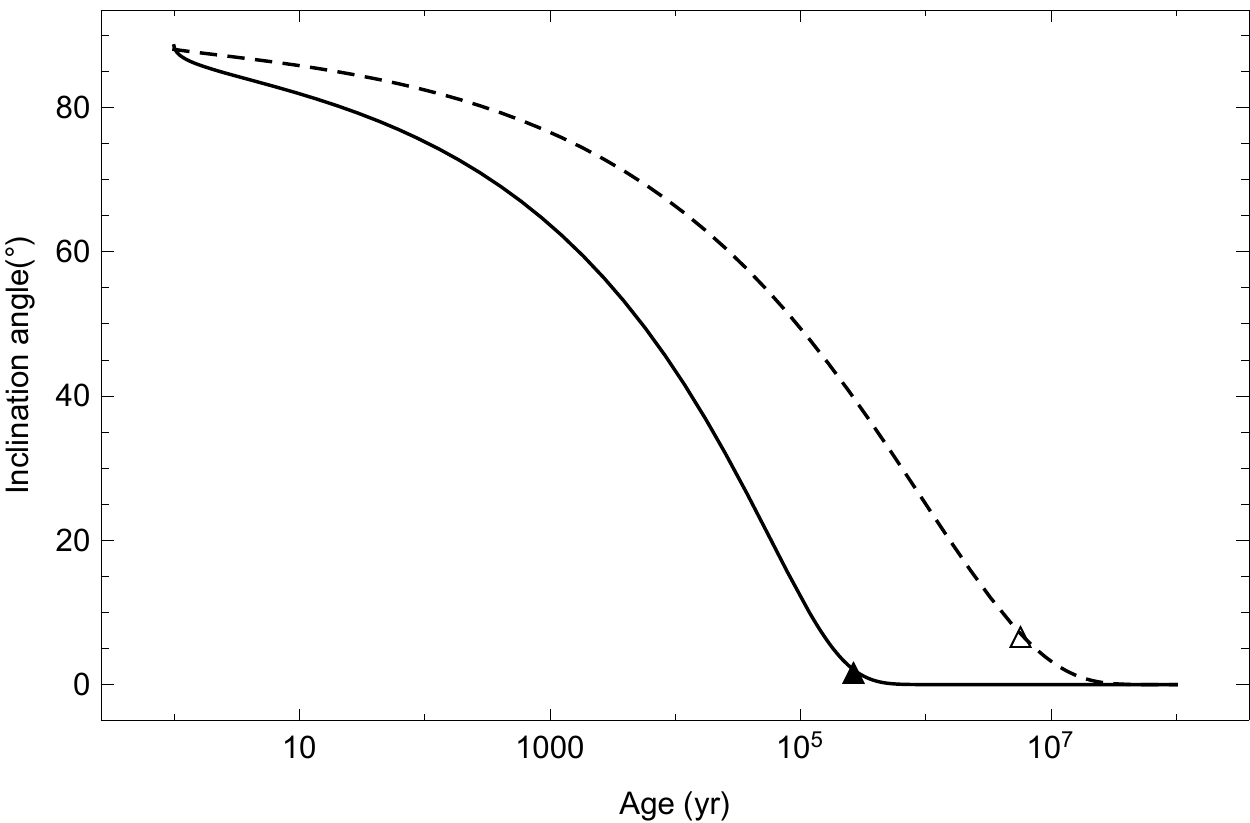}
\caption{The inclination angle evolution of PSR J0250+5854 in the case of magnetospheric evolution. The solid and dashed lines are calculations for ($B=7\times10^{14} \, \rm G$, $V_{\rm gap}=10^{13} \, \rm V$)  and ($B=10^{14} \, \rm G$, $V_{\rm gap}=10^{12} \, \rm V$), respectively. The solid triangle represents the present inclination angle of $2^{\circ}$ at age of $2.7\times10^5\, \rm yr$, and the hollow one is for $7^{\circ}$ and $5.4\times10^6\, \rm yr$. }
\label{fig:alpha}
\end{figure}

\begin{figure}[htp]
\plotone{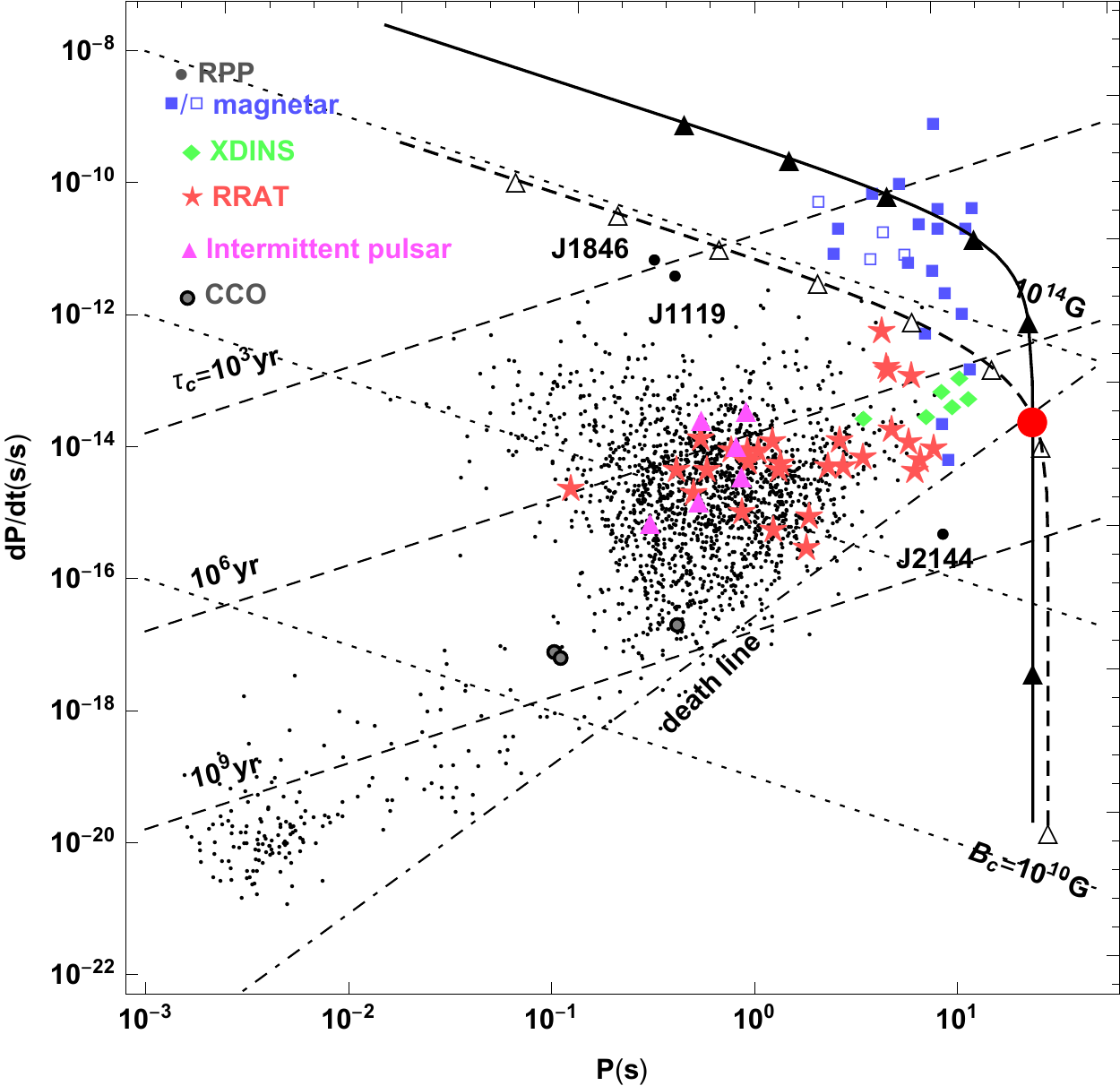}
\caption{Rotational evolutions of PSR J0250+5854 in the case of magnetospheric evolution in the $P$-$\dot{P}$ diagram. The solid evolution line is for ($B=7\times10^{14} \, \rm G$, $V_{\rm gap}=10^{13} \, \rm V$, $\alpha=2^{\circ}$) and the solid triangles represent the periods and period derivatives at  ages of $10\,\rm yr$, $100\,\rm yr $, $10^{3} \, \rm yr$, $10^{4} \, \rm yr $, $10^{5} \, \rm yr$, and $10^{6} \, \rm yr$. The dashed line is for ($B=10^{14} \, \rm G$, $V_{\rm gap}=10^{12} \, \rm V$, $\alpha=7^{\circ}$ and the hollow triangles represent the periods and period derivatives at  ages of $10\,\rm yr$, $100\,\rm yr $, $10^{3} \, \rm yr$, $10^{4} \, \rm yr $, $10^{5} \, \rm yr$, $10^{6} \, \rm yr$, $10^{7} \, \rm yr$ and $10^{8} \, \rm yr$, respectively. The large red point is the position of PSR J0250+5854. The black points are the positions of PSRs J1846-0258, J1119-6127 and J2144-3933, respectively.} The dot-dashed line is the pulsar death line based on curvature radiation from vacuum gap \citep{2000_Zhang_pulsar_death}. 
\label{fig:alphaevo}
\end{figure}

The coupled evolutions of inclination angle and rotation can be calculated by Equations  (\ref{equ:Omega3}) and (\ref{equ:alpha2}). Choosing the current period and magnetic field, the evolutionary paths for different assumed ages are the same, but with different beginning points. To get an initial period smaller than $20\, \rm ms$ \footnote{It is generally accepted  that the initial period of pulsar is about tens of milliseconds. An initial period of $17\, \rm ms$ was predicted for the Crab pulsar according to its truly recorded age \citep{2015_Lyne_Crab}.}, an age of $2.7\times10^5$ yr is assumed, and the corresponding initial period and inclination angle are $15\, \rm ms$ and $88.5^{\circ}$, respectively. The evolutions of inclination angle and rotation of PSR J0250+5854 are shown as solid lines in Figures \ref{fig:alpha} and \ref{fig:alphaevo}. The solid triangle in Figure \ref{fig:alpha} represents the inclination angle of $2^{\circ}$ at its age of $2.7\times10^5\, \rm yr$.  Meanwhile, the solid triangles in Figure \ref{fig:alphaevo} represent the periods and period derivatives at  ages of $10\,\rm yr$, $100\,\rm yr $, $10^{3} \, \rm yr$, $10^{4} \, \rm yr $, $10^{5} \, \rm yr$ and $10^{6} \, \rm yr$, respectively.   The death period and age are about $23.6\, \rm s$ and $2.1\times 10^{6} \,\rm yr$, respectively. 

Different acceleration models yield the potential drops in the order of $10^{12} \, \rm V$. A constant potential model of $3\times10^{12} \, \rm V$ was taken for approximate calculation \citep{1975ApJ...196...51R,2001_xu_wind}. Here, we discuss a relatively extreme case with $V_{\rm gap}=10^{12}\, \rm V$.  Similarly, a minimum magnetic field of $7.06\times 10^{13} \, \rm G$ and a maximum inclination angle of $36^{\circ}$ can be calculated by Equation (\ref{equ:Omega3}) and (\ref{equ:pdeath}).  Lower magnetic fields could be calculated by giving larger inclination angles.  However, the evolutionary trend almost keeps the same. Considering alignment of the inclination angle, as well as the pulsar's position on the $P$-$\dot{P}$ diagram, a small inclination angle is assumed for PSR J0250+5854. A magnetic field of $10^{14}\,\rm G$ will be calculated if an inclination angle of $7^{\circ}$ is assumed. Different ages used just result in different start points of the evolutionary line. Similarly, to get an initial period smaller than $20 \, \rm ms$, an age of $5.4\times10^6$ yr is assumed here. Coupled evolutions of the rotation and the inclination angle can be calculated by Equation (\ref{equ:Omega3}) and (\ref{equ:alpha2}). The initial inclination angle and rotational period are about $87.8^{\circ}$ and  $18\, \rm ms$, respectively. Evolutions of inclination angle and rotation of PSR J0250+5854 are shown as dashed lines in Figures \ref{fig:alpha} and \ref{fig:alphaevo}. The hollow triangle in Figure \ref{fig:alpha} represents its present inclination angle and age. The hollow triangles in Figure \ref{fig:alphaevo} are the periods and period derivatives at  ages of $10\,\rm yr$, $100\,\rm yr $, $10^{3} \, \rm yr$, $10^{4} \, \rm yr $, $10^{5} \, \rm yr$, $10^{6} \, \rm yr$, $10^{7} \, \rm yr$ and $10^{8} \, \rm yr$, respectively.  The death period and death age are about $28\, \rm s$ and $1.4\times10^8\,\rm yr$, respectively. 

As we can see from Figure \ref{fig:alpha}, the inclination angle decreases as the pulsar ages, which means that the magnetic axis and the rotational axis tend to align.  In the $P$-$\dot{P}$ diagram (Figure \ref{fig:alphaevo}), the pulsar first evolves to the right under the combined effect of particle winds and magneto-dipole radiation, and then turns down under the effect of pulsar ``death".  It is possible to predict that PSR J0250+5854 should be an old high magnetic field pulsar or magnetar, which is on the death edge.  
As the exhausting of the magnetospheric particles, its radio radiation will tend to stop. That may be why nulling pulses are observed \citep{2007_Zhang_RRATs_death,2015_Kou&Tong}. Besides, it may have a relatively small inclination angle due to the alignment of the rotational and magnetic axes \citep{2014_Philippov_3D,2017_Tong_alignment}. After the death point, the pulsar will only be spun-down by magneto-dipole radiation, and the effective magnetic field should be $B\sin\alpha$. From the calculation results, the magnetic field of PSR J0250+5854 is still high, in the order of $10^{14}\, \rm G$. Its true age is much smaller than its characteristic age. It is still possible to observe magnetar like emissions \citep{2011_Perna_magentar_bursting}. 

Besides, as we can see from evolution tracks in Figure \ref{fig:alphaevo}, the rotational period increases slowly after the death point because of the decrease in spin-down torque. The period derivative at death point is $\dot{P}_{\rm death}=5\times10^{-16}(P_{\rm death}/1\, \rm s)^3(V_{\rm gap}/10^{13} \, \rm V)^{2}\sin^{2}{\alpha}$ \citep{2006_Contopoulos_spindown,2012_Tong_SGR0418}.  For a pulsar with  $B=10^{14}\, \rm G$ and $V_{\rm gap}=10^{13}\, \rm V$, its maximum inclination angle is $3.4^{\circ}$, and its expected period and period derivative are $8.8\, \rm s$ and $1.2\times10^{-15} \, \rm s/s$, respectively. We could hence predict that for similar long-period pulsars, most of them should have a relatively small period derivative, i.e., $<10^{-15}\,\rm s/s$. 

\section{Rotational evolution  in the case of magnetic field decay}
In this case, rotational evolution of the pulsar under the effect of magnetic field decay will be discussed.  We also take the same assumption with $k_{0} \approx k_{1}\approx1$ in the spin-down torque, and pulsar rotational evolution should be \citep{2006_Spitkovsky_MHD} 
\begin{equation}
I\dot{\Omega}= -\frac{\mu^2}{c^3}\Omega^3(\sin^2\alpha+1)  =-\frac{B^2 R^{6}}{4c^3}\Omega^3(\sin^2\alpha+1).
\label{eqn:OmegaB}
\end{equation}

Since magnetars are generally taken as powered by high magnetic field, magnetic field decay is naturally and widely discussed \citep{1992_Goldreich_magentic_decay,1998_Heyl_magnetic_decay,2013_Vigan_thermal,2018_Popov_0250}. For an isolated neutron star, numerical calculations of magnetic field decay can be generally expressed in a simple power law with a decay index $\beta$, $\frac{\partial B}{\partial t} \propto -B^{1+\beta}$ \citep{2000_Colpi_clusting,2012_Dall_magnetic_decay}. The resulting changes are only quantitative for different forms of magnetic field decay. 
The magnetic energy may be dissipated by Hall/Ohmic diffusion in the stellar crust, and an analytical decay could be expressed as \citep{2008_Aguilera_magnetic_field,2012_Popov_magnetic_field}
\begin{equation}
B(t)=\frac{B_{0}\exp(-t/\tau_{\rm o})}{1+(\tau_{\rm o}/\tau_{\rm H})[1-\exp(-t/\tau_{\rm o})]}+B_{\rm fin},
\label{equ:Bdeacay}
\end{equation}
where $B_{0}$ is the initial field, $B_{\rm fin}$ is the relic field, $\tau_{\rm o}$ and $\tau_{\rm H}$ are the characteristic  time-scale of Ohmic and Hall decay, respectively. 

\subsection{Spin-down of  PSR J0250+5854}
\begin{figure}[htp]
\plotone{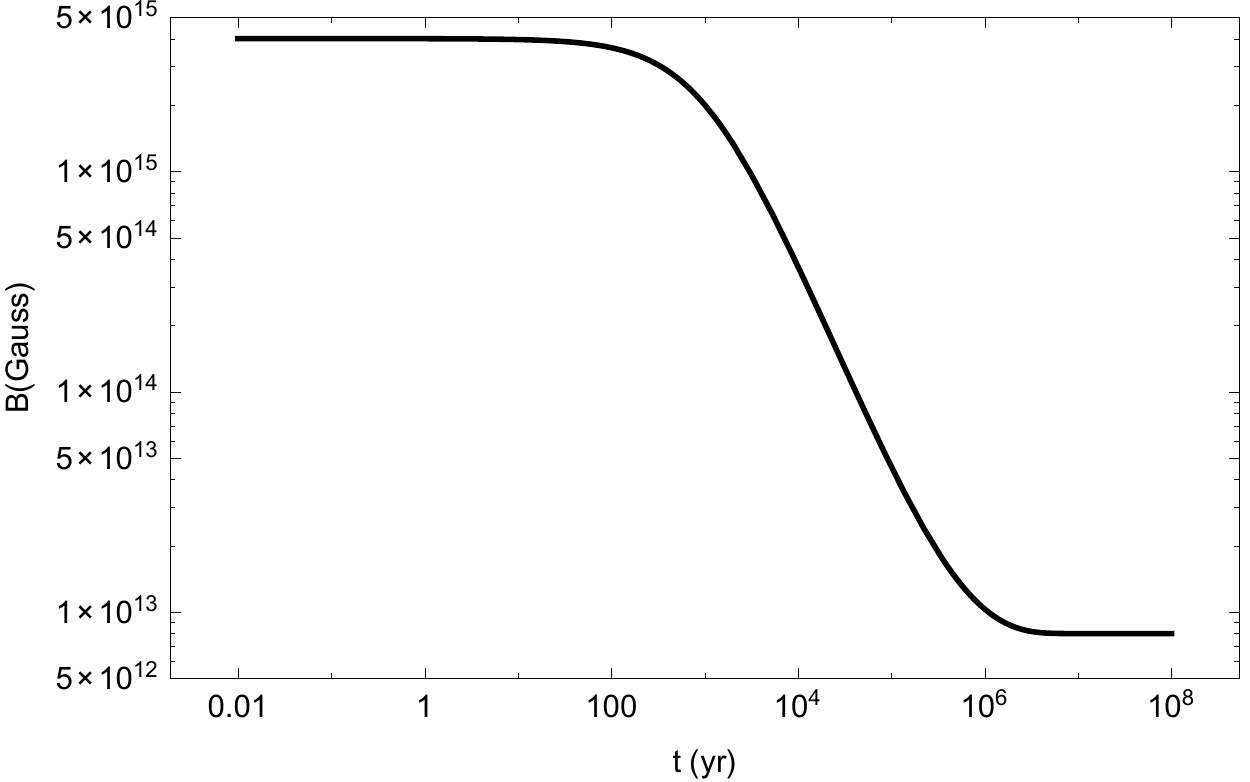}
\caption{The evolution of magnetic field with initial magnetic field of $4\times10^{15}\, \rm G$.\label{fig:Bdecay}}
\end{figure}

Assuming an initial magnetic field of $B_{0}=4 \times 10^{15} \rm G$, together with $\tau_{\rm o}=10^6 \rm yr$, $\tau_{\rm H}=10^3 \rm yr$ and $B_{\rm fin}=8\times10^{12} \, \rm G$, the magnetic field evolution can be calculated by Equation (\ref{equ:Bdeacay})\footnote{Compared with the magnetospheric evolution, the calculation manner is different for the magnetic field decay case. Because the evolutions of inclination angle and rotation are coupled in the magnetospheric evolution case, we assume an age for the pulsar to calculate its parameters and evolutionary path. However, the inclination angle is assumed to be stable in the magnetic field decay case. Given an initial magnetic field, its evolution could be calculated from the beginning.}, which is shown in Figure \ref{fig:Bdecay}. 

\begin{figure}[htp]
\plotone{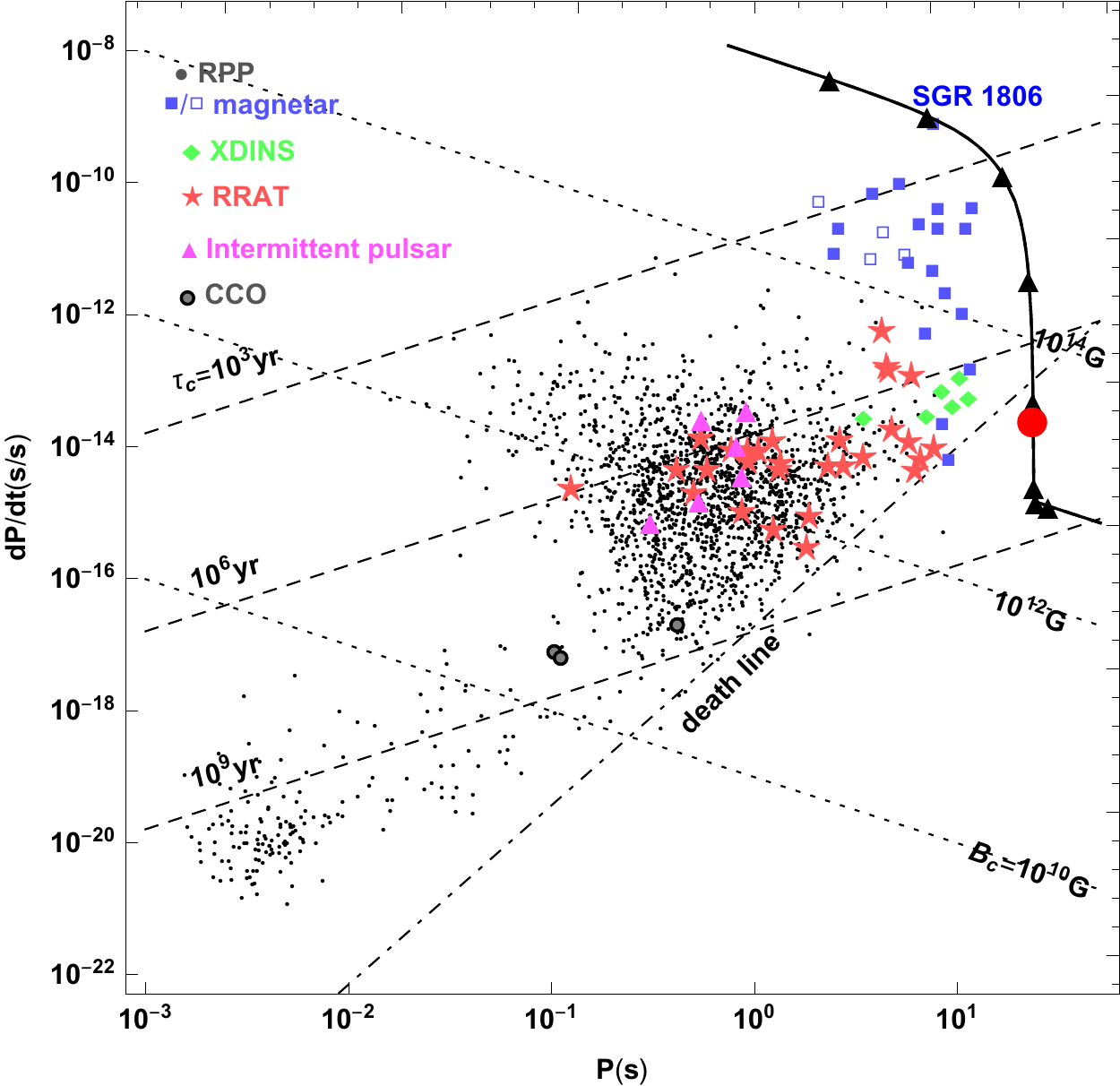}
\caption{Rotational evolution of PSR J0250+5854 in the case of magnetic field decay in the $P$-$\dot{P}$ diagram. The solid line is the evolution with initial magnetic field of $4\times 10^{15} \, \rm G$ and  period of $0.01\, \rm s$ together with a typical inclination angle of $45^{\circ}$. The solid triangles represent its periods and period derivatives at  ages of $10\,\rm yr$, $100\,\rm yr $, $10^{3} \, \rm yr$, $10^{4} \, \rm yr $, $10^{5} \, \rm yr$, $10^{6} \, \rm yr$, $10^{7} \, \rm yr$ and $10^{8} \, \rm yr$, respectively. The magnetar SGR 1806-20 is marked as solid blue square.
\label{fig:Bevo}}
\end{figure}

Given a typical inclination angle of $45^{\circ}$, together with an initial period of $0.01\, \rm s$, the rotational evolution of PSR J0250+5854 could be calculated by Equations (\ref{eqn:OmegaB}) and (\ref{equ:Bdeacay}), which is shown in Figure \ref{fig:Bevo}. The solid triangles represent its periods and period derivatives at  ages of $10\,\rm yr$, $100\,\rm yr $, $10^{3} \, \rm yr$, $10^{4} \, \rm yr $, $10^{5} \, \rm yr$, $10^{6} \, \rm yr$, $10^{7} \, \rm yr$ and $10^{8} \, \rm yr$, respectively.  The age and present magnetic field are about $1.4\times10^{5} \, \rm yr$ and $3.4\times 10^{13}\, \rm G$, respectively.  As we can see from Figure \ref{fig:Bdecay}, the magnetic field decays sharply at ages between $10^3$ and $10^6$ yr, which will result in a decrease in the spin-down torque. Correspondingly, the rotational evolution line quickly turns down after age about $10^3$ yr.  As shown in Figure \ref{fig:Bevo}, the evolution line passes through the region of SGR 1806-20, a so-called high magnetic field magnetar with dipole magnetic field about $4\times10^{15}\, \rm G$ at the pole (which is chosen as the initial magnetic field). It is possible to predict that the precursor of PSR J0250+5854 is a magnetar. Due to the decay of magnetic field, the magnetic-field powered magnetospheric activity will become weaker and tend to stop. From the calculation results, it may be hard to observe magnetar-like burst \citep{2011_Perna_magentar_bursting}. 

Besides, as the magnetic field tends to a constant value after $10^6$ yr, the pulsar will spin down under a constant spin-down torque, and $\dot{P}=4.0\times10^{-14}/P$ in case of $\alpha=45^{\circ}$ according to Equation (\ref{eqn:OmegaB}). In this case, for similar long-period pulsars, most of them should have relatively higher period derivatives, i.e., $\sim10^{-15}\,\rm s/s$. 

Several cases could be computed by varying the initial magnetic field $B_{\rm 0}$, the Hall time-scale $\tau_{\rm H}$, as well as the inclination angle $\alpha$. The evolution tracks are roughly the same, and the results only change quantitatively. The aim of this section is to probe the possible evolution path of the long period pulsar under the existing parameter space. Hence, a relatively high initial magnetic field is chosen here. 

\section{Discussions}
The pulsar will evolve along isodynamic lines under the pure magneto-dipole radiation model. Such a hypothesis is hard to explain observations of the pulsar. There should not be period clustering among pulsars in this case.  However, it would be different by considering physical braking torque.  \citet{2000_Colpi_clusting} discussed the period clustering of the anomalous X-ray pulsars (AXPs for short) by considering coupled evolution of thermal and magnetic field. The pulsar clustering could also be identified under these two spin-down models discussed in this paper, and, in terms of the period derivative. 

The pulsar death line is defined as a  threshold when pair production ceases, and it is model dependent \citep{1975ApJ...196...51R,1993_Chen_deathvelly,2000_Zhang_pulsar_death,2017_Zhou_death_line}.  For a specific pulsar, its death  depends on its magnetic field and the potential drop of the acceleration gap (Equation \ref{equ:pdeath}). Therefore, it is not strange to find pulsars under the death lines or outside the death valley. As discussed in \citet{2015_Kou&Tong}, the particle density may decrease as pulsar spinning down. Hence, for pulsars on the death edge,  their radio emissions tend to stop and they may be observed in abnormal ways, such as nulling,  intermittent pulsar, or ever rotational radio transients \citep{2007_Zhang_RRATs_death,2015_Karako_RRATs_review}. 

The evolution of  inclination angle is also widely studied, statistical studies show that long-period pulsars have relatively small inclination angle \citep{1988_Lyne_alpha_obs,1998_Tauris_alpha_obs}. The same conclusion are performed by a statistical study of the pulse width of radio pulsars with interpulse emission \citep{2011_Maciesiak_interpulse}. In a $3$-D magnetosphere, the magnetosphere may also generate a torque to align the rotational axis and magnetic axis.  In other words, the inclination angle tends to decrease \citep{2014_Philippov_3D,2017_Tong_alignment}. \citet{2012_Tong_SGR0418} had applied the spin-down model of \citet{2006_Contopoulos_spindown} to simulate the rotational evolution of SGR 0418+5729, a so-called low magnetic field magnetar. They pointed out that SGR 0418+5729 may be a normal magnetar but with a small inclination angle. From the point of view of inclination angle evolution, it is possible.  PSRs J1119-6127 and PSR J1846-0258 are two young pulsars with high magnetic field about $8\times10^{13}\, \rm G$ and $9.7\times10^{13} \, \rm G$ at the pole. Both of them are observed with magnetar-like bursts \citep{2008_Gavriil_1846,2016_1119_burst}. Besides, the radio emission of PSR J1119-6127 disappeared during the bursts, but reappeared about two weeks after the bursts \citep{2017_1119_raido}.  Connections between high-magnetic field  pulsar and magnetars are discussed \citep{2017_Archibald_1119}. The wind braking model is applied to explain their observations and simulate their long-term rotational evolution \citep{2016_Ou_Fluctuating_magnetosphere}. From the point of view of magnetosphere evolution, PSR J0250+5854 may be a high-magnetic field pulsar or magnetar on the death edge, it has a small inclination angle but a strong magnetic field. Its true age may be much smaller than the characteristic age.  Due to its relatively high-magnetic field, it may be burst active but with a relatively longer waiting time \citep{2011_Perna_magentar_bursting}. 

As the Hall and Ohmic diffusion are both temperature dependent, the magneto-thermal evolution is studied \citep{2009_Pons_thermal,2013_Vigan_thermal}.  For PSR J0250+5854, an age of $1.4\times10^{5} \, \rm yr$ is calculated with an initial magnetic field of $4\times10^{15} \, \rm G$. The expected thermal luminosity of  PSR J0250+5854 is about $10^{34} \, \rm erg/s$ via the magneto-thermal evolution models of \citet{2013_Vigan_thermal}. However, the Swift/XRT non-detection places an upper limit on luminosity of $1.5\times10^{32}\, \rm erg/s$ on $kT=85\, \rm eV$ for $N_{H} =9\times10^{21} \, \rm cm^{−2}$ \citep{2018_Vela}. 
One possibility is that the evolution of PSR J0250+5854 is dominated by magnetospheric evolution. However, because of the inner super dense matter of neutron star is still a mystery, it is also possible that PSR J0250+5854 has undergone a rapid cooling process \citep{2008_Yakovlev_cooling}.

In fact, both the effect of magnetospheric evolution and magnetic field decay may work on the pulsar, the dominant mechanism could also be identified by the period derivative and the distributions of long-period pulsars. In the case of magnetospheric evolution, the spin-down torque continues to decrease in the later stage, which will result in a continuous decrease in the period derivative (Figure \ref{fig:alphaevo}). Hence, a relatively low period derivative will be measured, for example, the radio pulsar PSR J2144-3933, with a period of  $8.5\,\rm s$ and a relatively low period derivative of $4.96\times10^{-16} \, \rm s/s$ \citep{1999_J2144}. However, in the case of magnetic field decay, the magnetic field tends to be constant at its late stage, and so is the spin-down torque. Pulsars will have a relatively high period derivative in this case. Clustering distribution of more similar long-period pulsars will help to distinguish these two cases. 

A fallback disk model is also commonly applied to explain the observations of pulsars and pulsar-like objects. Due to the significant increase in the period during the propeller phase,  the evolution of long-period neutron star could be understood by considering fallback disk torque \citep{2001_Menou_disk,2013_Li_fall_back_disk,2016_Tong_fall_back_disk}.  After the propeller phase, the neutron star tends to be in rotational equilibrium with the fallback disk at the equilibrium period of $P_{\rm eq}=2\pi(\frac{R_{\rm m}^3}{GM})^{1/2}$, where $G$ is the gravitational constant and $M$ is the mass of the neutron star. 

We also calculated the possible evolution of PSR J0250+5854 under the fallback disk model of \citet{2016_Tong_fall_back_disk}. Assuming that PSR J0250+5854 is spinning at the rotational equilibrium phase, a relatively large age about $10^{7} \, \rm yr$ could be estimated according to its period derivative: $\dot{P}\approx \frac{3\alpha P}{7t}$, where $\alpha$ is the power-law index of the mass accretion rate, and $\alpha=1.25$ is taken here. The model implies that the magnetic field of PSR J0250+5854 is about $10^{12}\, \rm G$, much smaller than its dipole magnetic field. Similarly, lower magnetic field about $10^{11} \, \rm G$ and larger age about $10^{8} \, \rm yr$ are expected for PSR J2144-3933, because of its low period derivative.  However, the typical lifetime of a fallback disk is generally taken as about a few thousand years \citep{2001_Menou_disk}. The model expected period derivatives would be much larger than the observed values for these two long-period pulsars. One possibility is that the fallback disk around them is not active and the pulsar is braked by magnetic-dipole radiation at the moment, the ages and magnetic fields of them are much closer to their characteristic ages and characteristic magnetic fields.  The pulsar will spin down along the isodynamic lines. Future measurements of the pulsar age and magnetic field would help to probe the possibility of braking under torques due to fallback disk for these long-period pulsars. 

In conclusion, PSR J0250+5854 may be an old high magnetic field pulsar or a magnetar. Considering the magnetospheric evolution, PSR J0250+5854 may be on the death edge, it may have a small inclination angle and a relatively high magnetic field, magnetar like emissions may be observed. However, in the case of magnetic field decay, the magnetic field powered magnetar like emission may be difficult to observe.  Besides, from the point of view of evolution studied, pulsars will have low period derivatives in the case of magnetospheric evolution, but  relatively higher period derivatives in the case of magnetic field decay.  Clustering distribution of more similar long-period pulsars will also help to distinguish these two cases. 

\textit{\small{This work is supported by the West Light Foundation of CAS (No. 2018-XBQNXZ-B-023), the National Key R$\&$D Program of China under grant number 2018YFA0404703 and the Open Project Program of the Key Laboratory of FAST, NAOC, Chinese Academy of Sciences. HT is supported by the NSFC (11773008). RXX is supported by the National Key R$\&$D Program of China (Grant No. 2017YFA0402602), the National Natural Science Foundation of China (Grant Nos. 11673002, and U1531243) and the Strategic Priority Research Program of Chinese Academy Sciences (Grant No. XDB23010200).  XZ is supported by the National Natural Science Foundation of China (Nos. 11873040, 11373006 and U1838108).  The FAST FELLOW-SHIP is supported by Special Funding for Advanced Users, budgeted and administrated by Center for Astronomical Mega-Science, Chinese Academy of Sciences (CAMS). }}




\bibliography{tex}




\end{document}